\documentstyle[preprint,eqsecnum,aps,prb]{revtex}

\tightenlines
\begin{document}
\draft
\title{Electron relaxation in  disordered gold films}
\author{B. I.  Belevtsev$^{*}$, Yu. F. Komnik, E. Yu. Beliayev}
\address{B. Verkin Institute for Low Temperature 
Physics \& Engineering, \\ 
Kharkov, 310164, Ukraine}
\date{\today}
\maketitle
\begin{abstract}

\par
The analysis of quantum corrections to magnetoconductivity of 
thin Au films responsible for by the effect of weak electron 
localization has made it possible to determine the temperature 
dependences of electron phase relaxation time in the temperature 
range 0.5--50 K for different degrees of crystal lattice disorder. 
The disorder was enhanced by irradiating the films in vacuum with 
3.5 keV Ar ions. The experimental data clearly demonstrate that the 
contribution of electron-electron interaction to electron phase
relaxation increases with disorder and support the theoretical 
prediction that the frequency of electron-phonon scattering tends 
to diminish upon a decrease in electron mean free path. It is found 
that the spin-orbit scattering rate decreases with disorder. In our 
opinion, such unusual behavior can take place for thin films at 
decreasing the electron mean free path provided, that the surface 
electron scattering contributes significantly to the total spin-orbit 
scattering.
\end{abstract}

\narrowtext

\section{Introduction}

\label{sec:Int}

\par
The problem of crystal lattice disorder effect upon the processes of
inelastic and phase relaxation of electrons in metal systems continues 
to be topical. Theory \cite{1,2} demonstrates that the electron-electron
interaction increases with disordering and the behavior of temperature
dependence of electron-electron relaxation time $\tau_{ee}$ depends on
system dimensionality. Thus, for a two-dimensional system $\tau_{ee}^{-1} 
\propto T$, not to $T^{2}$ , as it follows from the Fermi-liquid theory 
for pure metals. Electron relaxation time due to inelastic electron-phonon 
collisions, $\tau_{ep}$, in pure metals is characterized by the dependence 
$\tau_{ep}^{-1} \propto T^{3}$ which is also true for weak disorder 
\cite{3} with $q_{T}l \gg 1$ ("pure limit") where $q_{T}$ is the thermal 
phonon wave vector and $l$ is the electron mean free path. In the case of 
strong disorder with $q_{T}l \ll 1$ ("dirty limit"), the temperature 
dependence of $\tau_{ep}^{-1}$ should take the form $\tau_{ep}^{-1} 
\propto lT^{4}$ as reported in \cite{4,5,6}. The increase in exponent 
on $T$ and the unexpected dependence on $l$ (the frequency of 
electron-phonon scattering decreases with $l$) can be attributed to the 
reduction in the electron-phonon interaction on the crossover to a dirty 
limit. The processes of inelastic electron scattering by oscillating 
impurities that result in the dependence $\tau_{ep}^{-1} \propto 
l^{-1}T^{2}$ are considered in Ref.\ \onlinecite{7}. According to Ref.\ 
\onlinecite{4} this contribution makes itself evident under electron 
scattering by transverse phonons at $q_{T}l > 1$. However, as shown in 
Ref.\ \onlinecite{6}, the contribution of this mechanism is cancelled 
out in dirty limit and is missing in the pure one \cite{4}.

\par
Hence the pure-dirty limit transition must involve the change of the 
dependence $\tau_{ep}^{-1} \propto T^{3}$ to $\tau_{ep}^{-1} \propto 
lT^{4}$. Therefore, for disordered metals both the behavior of 
temperature dependence of $\tau_{ep}$ and the effect of electron mean 
free path on $\tau_{ep}$ are currently central problems.

\par
In disordered metals the interference of electron wave functions at low 
temperatures results in quantum corrections to conductivity that may be 
attributed to the effects of weak localization (WL) \cite{8,9,10,11} and 
electron-electron interaction (EEI) \cite{1,2}. The values of these 
corrections are directly related to the characteristic electron relaxation 
times and electron-electron interaction constants. This permits the phase 
relaxation time of electron wave function, $\tau_{\varphi}$, (in essence, 
the inelastic relaxation time) and the spin relaxation time of electrons 
due to spin-orbit ($\tau_{so}$) and spin-spin ($\tau_{s}$) interaction to 
be determined from the analysis of the experimental temperature and 
magnetic field dependences of conductivity, $\sigma (T,H)$.

\par
It was found \cite{12} that the temperature dependence of $\tau_{\varphi}$ 
determined from the quantum corrections to magnetoconductivity of Au films 
could be described by the relation $\tau_{\varphi}^{-1} \propto T^{2}$ at 
$T > 4.5$~K, the latter being associated with the electron-phonon relaxation 
processes. More recently, the same result was obtained in Refs.\ 
\onlinecite{13,14,15}. According to Ref.\ \onlinecite{14}, the contribution 
of electron-phonon relaxation separated from $\tau_{\varphi}(T)$ for an Au 
film of $R_{\Box} \approx 10^{3}$~$\Omega$ ($R_{\Box}$ is sheet resistance) 
could be described by the relation $\tau_{ep}^{-1} = A_{p}T^{2}$, where 
$A_{p} = 1.42\times 10^{9}$~s$^{-1}$K$^{-2}$. However, the overheating 
experiments at ultralow temperatures (0.03--1~K) described in Ref.\ 
\onlinecite{16} for Au films of $R_{\Box} \simeq 30-100$~$\Omega$ resulted 
in an expression $\tau_{ep}^{-1} = A_{p}T^{p}$, where $p = 2.7$--$2.9$ and 
$A_{p}  = 10^{8}$~s$^{-1}$K$^{-3}$. A similar relation $\tau_{ep}^{-1} = 
1.3\times 10^{8}T^{3}$~s$^{-1}$ is consistent with the data on temperature 
dependence of $\tau_{\varphi}$  obtained in Ref.\ \onlinecite{17} for Au 
films of $R_{\Box} \simeq 7$~$\Omega$ in  the temperature range 0.05--10 K.
Thus, for Au films $\tau_{\varphi}^{-1}$ and/or $\tau_{ep}^{-1}$ have a 
$T^{2}$ dependence in some cases and a $T^{3}$ dependence in other cases. 
The $\tau_{ep}^{-1} \propto T^{2}$ dependence was also found for Nb films 
from the effect of electron overheating in the temperature range 1.5--10K 
\cite{18}. The $\tau_{ep}^{-1} \propto T^{3}$ dependence was as a 
rule observed for pure films of other metals. For example, in Ref.\
\onlinecite{19} a temperature dependence of the inelastic time was
obtained for pure Al films from weak localization. From this dependence
the contribution of electron-phonon scattering was separated, which
agreed with $\tau_{ep}^{-1} = A_{p}T^{3}$ calculated from the real band
structure of Al for electron scattering by longitudinal phonons. A
similar result was also obtained for Al films later \cite{20,21}. For
pure Cu films the dependence $\tau_{ep}^{-1} \propto T^{3}$ for the time
of electron-phonon energy relaxation at ultralow temperatures (10-300~
mK) was obtained from the effect of electron overheating \cite{22}.

\par
The effects of electron mean free path and film thickness on 
electron phase relaxation time in Au, Ag and Mg films were studied 
in Ref.\ \onlinecite{13}. The electron mean free path, $l$, was varied 
by annealing of quench-condensed films by stages. The experiments did not 
reveal any dependence of $\tau_{\varphi}$  on electron mean free path. 
The same result was obtained for the films of some other metals, for 
instance, Sb \cite{23} and Bi \cite{24}. However, for Nb films the
dependence upon the mean free path of the $\tau_{ep}^{-1} \propto
lT^{2}$-type was found in Ref.\ \onlinecite{18}.
In our opinion, the results given in Refs.\ \onlinecite{23,24} may be 
attributed to a "pure limit". A contradictory situation appears in Ref.\
\onlinecite{18}: the $\tau_{ep}(l)$ dependence points to the "dirty"
limit, though the exponent on T is much below 4.

\par
This paper is concerned with the effect of disordering on 
characteristic electron relaxation times in thin Au films 
($\tau_{\varphi}, \tau_{ep}, \tau_{so}$) estimated by analyzing 
the quantum corrections to conductivity. The disorder was enhanced 
by irradiating the films with 3.5 keV Ar  ions, resulting in radiation 
damage of the crystal lattice. Variation in the fluence enabled us to 
prepare a series of samples of much the same thickness ($\simeq 10$~nm) 
with the values of $R_{\Box}$ between 6 and 500~$\Omega$. This 
permitted the previous data on manifestation of quantum interference 
effects in Au films \cite{12,13,14,15,16,17} to be complemented and 
new and, in some cases, unexpected features of disorder influence on 
phase and spin relaxation of electrons to be revealed.

\section{Samples}

\par
The Au films with thickness $\simeq$~10~nm were prepared by condensing 
a metal molecular beam at a rate 0.05~nm$\cdot$s$^{-1}$  at pressure 
of residual gases $P = 10^{-4}$~Pa. 99.99 \% purity gold was evaporated 
out of a Mo boat onto a sapphire single-crystal substrate (for measuring 
film conductivity) and a NaCl single-crystal cleft (for TEM study). 
Both substrates were precoated with a SiO sublayer ($\simeq$~15 nm 
thick) to ensure identical structural and topological parameters of 
the deposited films. To enhance the structure disordering, the films were 
bombarded in vacuum with argon ions of energy 0.5-3.5 keV. The ion 
fluences $\Phi$ varied within $1\times 10^{15}$  and 
$7\times 10^{15}$~cm$^{-2}$.

\par 
With this irradiation dose the decrease in the film thickness was
negligible (10-20 \% with the highest dose), but with a large dose
holes with the sizes of $\leq 5$ nm appeared in the film. The total 
surface fraction of the holes under the highest irradiation dose was 
$\leq 20 \%$. The electron microscopic results for the films are 
reported in Ref.\ \onlinecite{25}.

\par
As the irradiation dose was increased, the film resistance became
tens of times higher against its initial value. We suggest that the 
irradiation induced increase in the film resistance is substantially 
associated with radiation defects generated in the lattice. For the 
interference effects under consideration, which result from frequently 
occurring elastic processes of electron scattering, the nature of 
scatterers is of no importance. What counts is that the electron 
motion be of diffusion nature because of small electron mean free paths.

\par
The effects of WL and EEI in the films studied manifested 
themselves in the dependences of resistance $R$ on perpendicular 
magnetic field $H$ (varied up to 20 kOe) and temperature (varied 
between 0.5 and 50 K). The resistance measurements were carried out 
by using a standard four-probe dc technique. The values of current 
(5--20 $\mu$A) were chosen in such a way as to prevent, where possible, 
the influence of Joule heating. The relative error of the 
resistance measurements was $\leq 10^{-5}$. The film area under 
measurement was 2 mm $\times$ 0.1 mm in size. The samples were places into 
a vacuum cryostat with a superconducting solenoid where the 
sapphire substrate was in a good thermal contact with a copper 
vessel filled with $^{3}$He to produce temperatures below 1.5 K.

\par
Preliminary results on the problem discussed are reported in Ref.\ 
\onlinecite{25,26}. Our earlier data and the new experimental results 
of this study were computer-processed using an advanced program to fit
experiment results and theoretical formulas. A very good description was
obtained, which made conclusions more reliable and convincing.

\section{Calculations}

\par                 
To analyze the temperature and magnetic field dependences of 
conductivity $\sigma$, well-known expressions for quantum corrections 
$\Delta \sigma (T,H)$ associated with WL and EEI effects in 
two-dimensional systems \cite{10,11} have been used. For the films 
studied, the two-dimensional conditions of manifestation of quantum 
interference effects are fulfilled: $L < L_{\varphi}, L_{T}$, where $L$ 
is the film thickness, $L_{\varphi}$ = $(D\tau_{\varphi})^{1/2}$ is the 
diffusion length of phase relaxation, $L_{T} = (\hbar D/kT)^{1/2}$ is 
the thermal coherence length in a normal metal, and $D$ is the electron 
diffusion coefficient.

\par
The contribution of WL and EEI to the temperature dependence 
of conductivity in a zero magnetic field is of the form:

\begin{equation}
\Delta \sigma (T) = \frac{e^{2}}{2\pi^{2}\hbar}\left\{-\left 
[\frac{3}{2}\ln \frac{\tau_{\varphi}^{\ast}}{\tau} - 
\frac{1}{2}\ln \frac{\tau_{\varphi}}{\tau}\right ]
+ \lambda_{T}^{D}\ln \frac{kT\tau}{\hbar} \right\}
\label{one}
\end{equation}

where $\tau$ is the elastic electron relaxation time, 
$\tau_{\varphi}^{-1} = \tau_{\varphi 0}^{-1} + 2\tau_{s}^{-1}$; 
$(\tau_{\varphi}^{\ast})^{-1}  = \tau_{\varphi 0}^{-1} + 
(4/3)\tau_{so}^{-1} + (2/3)\tau_{s}^{-1}$, $\tau_{\varphi 0}$ is the 
phase relaxation time due to inelastic scattering, $\lambda_{T}^{D}$ 
is the interaction constant in the diffusive channel. The latter can
be written in terms of the universal constant $F$ - the angle-averaged 
amplitude of electron interaction with a small transferred momentum. 
$\lambda_{T}^{D} = 1 - (3/4)F$ in weak magnetic fields and 
$\lambda_{T}^{D} = 1 - (1/4)F$ in strong ones. For typical metals $F$ 
is close to zero. Below we use the presentation $\tau_{\varphi 0}^{-1} 
\propto T^{p}$, where $p$ is the exponent which depends on the 
mechanism of inelastic scattering. The first term in Eq.\ (\ref{one}) 
corresponds to WL effects, while the second one to EEI. The
variation of conductivity in an arbitrary temperature range from
$T_{1}$ to $T_{2}$ due to WL and EEI effects is 

\begin{equation}
\sigma (T_{1}) - \sigma (T_{2}) = - \frac{e^{2}}{2\pi^{2}\hbar} 
\left\{ \left [ 
\frac{3}{2} \ln 
\frac{\tau_{\varphi}^{\ast}(T_{1})}{\tau_{\varphi}^{\ast}(T_{2})} 
- 
\frac{1}{2} \ln 
\frac{\tau_{\varphi}(T_{1})}{\tau_{\varphi}(T_{2})}
\right]
+ \lambda_{T}^{D} \ln \frac {T_{2}} {T_{1}} 
\right\}
\label{two}
\end{equation}

\par  
Below we use $ -\Delta \sigma = (\Delta R)/(RR_{\Box})$, which is 
true for small corrections.

\par
The variation of conductivity in a perpendicular magnetic field 
associated with the WL effect in two-dimensional system can be 
given by the expression:

\begin{equation}
\Delta \sigma (H) = \frac{e^{2}}{2\pi^{2}\hbar}\left\{
\frac{3}{2}f_{2}\left (\frac{4eDH\tau_{\varphi}^{\ast}}{\hbar c}\right ) - 
\frac{1}{2}f_{2}\left (\frac{4eDH\tau_{\varphi}}{\hbar c} \right )
\right\}
\label{three}
\end{equation}

where $f_{2}(x) = \ln (x) + \Psi (1/2 + 1/x)$, $\Psi$ is digamma function. 
For the films studied, the contribution of EEI to magnetoresistance (MR) is 
negligible, so we do not cite corresponding expressions.

\section{Results and discussion}

\subsection{$\Delta \sigma (T,H)$ dependences}

\par                     
As an example, Fig.~\ref{Fig.1} shows the experimental
$\Delta\sigma (H)$ dependences obtained at different
temperatures(up to 50 K) for one of the samples. It can be seen
that the MR is positive which is the case of a strong spin-orbit
interaction ($\tau_{so} \ll \tau_{\varphi 0}(T)$). However, the
curves $\Delta\sigma (H)$ are nonmonotonic: for nonirradiated
samples a well-defined maximum can be seen. The shape of $\Delta
\sigma (H)$ curves reflects, according to Eq.\ (\ref{three}), a 
competition of inelastic and spin-orbit electron scattering
contributions.

\par
Using computer fitting procedure according to Eq.\ (\ref{three})
the characteristic values $D\tau_{\varphi}$ and $D\tau_{so}$ can be
found from the experimental data. In order to determine
$\tau_{\varphi}$ and $\tau_{so}$, the diffusion coefficient $D =
(1/3)v_{F}l$ has been used, where Fermy velocity $v_{F}$ is found
from the free-electron model and $l$ from the relation $\rho l =
8.39\times 10^{-12}$~$\Omega \times $cm$^{2}$ for Au \cite{15}
(where $\rho$ is the film resistivity, calculated, taking into
account the electron-microscopic data for the topological structure
of the films). The experimental magnetoresistive data are shown in
Fig.~\ref{Fig.1} by dots, while solid curves represent the
calculated $\Delta \sigma (H)$ dependences using Eq.\ (\ref{three})
with $D\tau_{\varphi}$ and $D\tau_{so}$ as fit parameters. As can
be seen from Fig.~\ref{Fig.1} the fitting accuracy was high enough.

\par
Fig.~\ref{Fig.2} shows a typical $D\tau_{\varphi}(T)$ dependence.
The $D\tau_{\varphi}(T)$ dependences characterize the temperature
variation of $\tau_{\varphi}$ since D may be assumed to be
temperature-independent. It can be seen that the character of this
dependence changes with decreasing temperature. Within the
temperature range 10--50 K, the $\tau_{\varphi} \propto T^{-p}$
dependence is revealed, where $p \approx 2$, in the range 3--10 K,
$\tau_{\varphi} \propto T^{-1}$ dependence has been observed for
high-resistive samples and at lower temperatures ($T < 2$~K)
$D\tau_{\varphi}$ goes to some constant value.

\par
The saturation of $\tau_{\varphi}(T)$ dependence at very low
temperatures may be attributed to several factors. One of them is
the influence of electron overheating caused by the measuring
current.  We however believe, that this factor is hardly probable
since the used measuring currents were quite small. Besides,
electron overheating does not lead to saturation and does not
change the slope of the dependence $\tau_{\varphi}(T)$ though the
$\tau_{\varphi}(T)$ values differ from the true ones. It is more
likely that saturation of $\tau_{\varphi}(T)$ is due to spin
electron scattering. In this case the saturation occures when spin
scattering by magnetic impurities at very low temperatures starts
predominating over the processes of electron-electron and
electron-phonon scattering, and the characteristic time of spin
scattering is temperature-independent.

\par
A new explanation proposed in Ref.\ \onlinecite{27} attributes the
low-temperature saturation of $\tau_{\varphi}(T)$ dependences in
one-dimensional (1D) and two-dimensional (2D) electron systems to
the influence of zero-point fluctuations of phase coherent
electrons, which leads to intrinsic decoherence of electrons. This
model accounts not only for its authors' results taken on narrow
(quasi-1D) Au films, but it also explains the $\tau_{\varphi}(T)$
saturation in 1D and 2D metallic and semiconducting samples,
observed by other reseachers.  Besides, the characteristic
saturation time $\tau_{0}$, calculated within the model of
fluctuation decoherence, agrees well with the experiment. As the
resistance of the samples increases (and hence the diffusion
coefficient of the electrons decreases), the time $\tau_{0}$
becomes shorter. This regularity is observed, but only for
high-ohmic samples with $R_{\Box} > 60 \Omega$, and the changes are
considerably smaller than in Ref.\ \onlinecite{27}.  In our samples
the saturation occures in comparatively low-Ohmic samples too. This
implies that the model used in Ref.\ \onlinecite{27} cannot be the
only explanation of saturation of $\tau_{\varphi}(T)$ dependence at
low temperatures.

\par
Above 4 K, where the factors responsible for the low temperature 
saturation of the dependence $\tau_{\varphi}(T)$ turn out to be 
insignificant, this dependence can be approximated by 

\begin{equation}
\tau_{\varphi}^{-1}(T) = \tau_{ee}^{-1}(T) + \tau_{ep}^{-1}(T)
\label{four}
\end{equation}

in which the electron-electron relaxation rate is numerically
determined by the expression in Ref.\ \onlinecite{10}:

\begin{equation}
\tau_{ee}^{-1}(T) = 
\frac{\pi kT}{\hbar}\: \frac{e^{2}R_{\Box}}{2\pi^{2}\hbar}\: \ln 
\left (\frac{\pi \hbar}{e^{2}R_{\Box}}\right )
\label{five}
\end{equation}

\par
Thus, at temperatures below 2 K, the character of temperature
dependence of $\tau_{\varphi}(T)$ depends on the times $\tau_{0}$
and $\tau_{s}$. At higher temperatures the shape of the curve
$\tau_{\varphi}(T)$ is first dictated by the predominating
electron-electron (ee) relaxation processes, and for the
temperatures $T > 20$~K - by the processes of electron-phonon (ep)
relaxation.

\par
To separate the contributions of the e-e and e-p relaxation to
$\tau_{\varphi}$, we used the following procedure. Assume that
$\tau_{\varphi}^{-1}(T)$ is described by the expression 

\begin{equation}
\tau_{\varphi}^{-1}(T) = AT+A_{p}T^{p}
\label{six a}
\end{equation}

or 

\begin{equation}
\left[ \tau_{\varphi}(T) \right]^{-1} = A + A_{p} T^{p-1}
\label{six b}
\end{equation}

\par
It is not reliable to estimate the parameters $A$, $A_{p}$ and $p$ 
from the fitting of experimental results to Eq.\ (\ref{six a}),
because the error is about 50 \% in this case. Our experimental
dependences were plotted in the coordinates $1/(\tau_{\varphi}T)$
vs. $T$. For high-Ohmic samples ($R_{\Box}>60 \Omega$), a horizontal
part appears in the curve in the region of dominating 
electron-electron relaxation (see insert in Fig.~\ref{Fig.2}) from
which the coefficient A can be estimated. The corresponding
$\tau_{ee}$ values turn out to be close to those, calculated from 
Eq.\ (\ref{five}). It is also found in earlier studies Ref.\
\onlinecite{14,16} that the ratio of calculated and experimental
$\tau_{ee}$-values for Au films is close to 1 ($\approx$
1.05--1.2). This permits us to take $\tau_{ee}$ equal to the value,
given by Eq.\ (\ref{five}). Then, $A_{p}$, and $p$ can be found with
good accuracy (the error is below 15\%). 

\par
For low-Ohmic samples $(R_{\Box} < 60 \Omega)$ the
electron-electron relaxation is not essential since the inequality
$\tau_{ee} \gg \tau_{ep}$ holds in the whole temperature range down
to $\approx 5$~K. Spin-spin scattering or electron decoherence is
observed below this temperature. For these samples no horizontal
part is seen in the $1/(\tau_{\varphi}T)$ vs. $T$ dependences and at
$T > 5$ K the $\tau_{\varphi}(T)$ dependence actually
characterizes electron-phonon relaxation. For low-Ohmic samples it
can be described by the $\tau_{ep}^{-1} \propto T^{2}$ dependence.
  
\par
We have found that at enhancing disorder and growing $R_{\Box}$ the
$\tau_{\varphi}$ behavior is different in different temperature
ranges: $\tau_{\varphi}$ increases in the region of predominating
electron-phonon scattering and goes down in the region, where the
e-e scattering makes itself evident. (Fig.~\ref{Fig.3}). The figure
is a vivid demonstration of the general statement: as disorder 
increases, the electron-phonon interaction grows weaker, while 
electron-electron interaction is enhanced.

\par 
The reliability of obtained $\tau_{\varphi}$ and $\tau_{so}$ values
can be illustrated while comparing the temperature-resistance
dependence calculated using $\tau_{\varphi}(T)$ and $\tau_{so}$
derived from magnetoresistance with the experimental $R(T)$ data.
The temperature variation of resistance in one of the samples is
shown in Fig.~\ref{Fig.4}. At $T > 20$~K we observe the dependence
$\Delta R/R \propto T^{2}$, which can be safely attributed
\cite{26} to electron-phonon-impurity interference \cite{28}. 
(see also the experimental results in Ref.\ \onlinecite{17,29}).
At lower temperatures we see (Fig.~\ref{Fig.4}) a horizontal part
and then a weak growth of resistance as the temperature goes down.
This behavior of the resistance can be explained well in terms of
the WL and EEI effects. For the strong spin-orbit interaction
($\tau_{so} \ll \tau_{\varphi 0}(T)$ and $\tau_{s} \gg \tau_{so},
\tau_{\varphi 0}$) Eq.\ (\ref{one}) can be written as  

\par
\begin{equation}
\frac{\Delta R}{R} = 
- a_{T}\, \frac{e^{2}R_{\Box}}{2\pi^{2}\hbar}\: \ln T + const, 
\label{seven}
\end{equation}

where $a_{T} = 1 - p/2$. For the electron-phonon interaction processes 
$p \approx 2$ and hence $a_{T} \approx 0$. If the electron-electron 
interaction predominates, $p = 1$ and consequently $a_{T} = 1/2$. 
Fig.~\ref{Fig.4} shows exactly this type of dependence $\Delta R(T)$. 
The open circles indicate the resistance calculated by Eq.\ (\ref{two}) 
using $\tau_{\varphi}(T)$ and $\tau_{so}$  found from 
magnetoresistance, while full circles represent the experimental 
$R(T)$ values. The agreement of calculations and the experiment is 
quite good.

\par 
Thus, the temperature behavior of quantum corrections $\Delta
\sigma (T)$ for Au films studied is consistent with what is 
expected theoretically for 2D systems which are far-away from the
percolation threshold. In the case of percolation effects, the
coefficient $a_{T}$ has to be less than the theoretical value.
Beside this, $\approx 1.5$-time decrease of $p$ in the expression
$\tau_{\varphi}^{-1} \propto T^{p}$ is typical of percolation
systems \cite{15}. This effect has not been found in the films
studied. 

\par
With the separated contribution of the e-e relaxation to
$\tau_{\varphi}$ it turns out, that for high-resistance samples
($R_{\Box} > 60$~$\Omega$) the temperature dependence of
$\tau_{ep}$ can be described by the function $\tau_{ep}^{-1}
\propto A_{p}T^{p}$ where $p$ is slightly above 2 and increases 
monotonically with $R_{\Box}$ (see the Table). The behavior of the 
immediate dependence of $\tau_{ep}$ on $l$ can be demonstrated if 
we plot the value of $\tau_{ep}$ versus $l$ for a certain 
temperature (Fig.~\ref{Fig.5}, $T = 20$~K). It turns out that some 
samples do not reveal any dependence of $\tau_{ep}$ on $l$ whereas 
the remaining samples feature a strong dependence of the form 
$\tau_{ep}^{-1} \propto l$. 

\par
The increase in the exponent $p$ with enhancing disorder (see 
the Table) and the existence of the above-mentioned 
relationship between $\tau_{ep}$ and $l$ ($\tau_{ep}^{-1} \propto l$)
suggest that as the films are disordered, there occurs a pure 
$(q_{T}l > 1)$ -- dirty $(q_{T}l < 1)$ limit transition. Indeed, the 
condition $q_{T}l \simeq 1$ defines a particular temperature 
$T_{tr} = \hbar s_{l,t}/kl$ ($s_{l,t}$  is the velocity of 
phonons of different polarization) below which a dirty limit case 
occurs, and this temperature increases with decreasing $l$ (see the 
Table). In those cases where $T_{tr}$ is below 
the temperature range within which the dependence 
$\tau_{\varphi}^{-1} \propto T^{2}$  makes itself evident, 
i.e. $T_{tr} < 10$~K, there is no influence of electron mean free path 
on $\tau_{ep}$. However when $T_{tr}$ is much higher than 10 K, an 
increase in $p$ is observed and the relation between $\tau_{ep}$
and $l$ appears.

\par
In view of the above behavior, the discrepancy between the 
$p$-values for Au films given in Refs.\ \onlinecite{12,13,14,15,16,17}
can be explained. The subjects of investigation in 
Refs.\ \onlinecite{12,13,15} correspond to a pure limit 
case, where $p$ is found to be equal to 2; the data in 
Ref.\ \onlinecite{16} are true for a dirty limit case (0.03--1.0 K) 
where $p$ took a value close to 3.

\par
The problem of reduction of the exponent $p$ as opposed to the 
theoretical predictions is much more difficult to interpret. As 
mentioned in Sec.\ \ref{sec:Int}, according to Refs.\ \onlinecite{4,5,6}
the bulk disordered samples must reveal dependences 
$\tau_{ep}^{-1} \propto lT^{4}$ for $q_{T}l < 1$ and 
$\tau_{ep}^{-1} \propto T^{3}$ for $q_{T}l > 1$. Since $\tau_{ep}$ 
is given by the Eliashberg function $\alpha^{2}(\omega)F(\omega)$ 
within the frequency range corresponding to the energy 
of thermal phonons \cite{30}:

\begin{equation}
\tau_{ep}^{-1} = 
4\pi \int \! d\omega 
\frac{\alpha^{2}(\omega)F(\omega)}{\sinh (\hbar \omega/kT)}, 
\label{eight}
\end{equation}

the reduction of $p$ from 3 to 2 within the pure limit corresponds to 
the occurrence of a linear, rather than quadratic, dependence of 
the Eliashberg function on $\omega$ at low frequencies.

\par
The change in the Eliashberg function form at low phonon 
frequencies may be accounted for by the modification of the phonon 
spectrum in thin films. In this case for the 2D phonon 
spectrum a dependence $\tau_{ep}^{-1} \propto lT^{3}$ is assumed 
to be observed in the dirty limit \cite{31,32} and 
$\tau_{ep}^{-1} \propto T^{2}$ in the pure one (rigorous theory 
is unavailable). Assuming two-dimensionality of the phonon spectrum 
allowed the authors of Ref.\ \onlinecite{16} to fit the calculated data 
on $\tau_{ep}$ with the experimental ones, and the process of heat escape to 
the environment at electron overheating was successfully 
interpreted in Ref.\ \onlinecite{33} on the same ground. 

\par
Many authors believe that the 2D phonon spectrum in a thin film is
due to the quantum size effect for the phonons when the phonon
wavelength $\lambda = 2\pi /q$ is comparable to the film thickness.
This problem is consequently discussed for free standing film. For
$L < \lambda$ there exist flexural waves with the quadratic
dispersion relations $\omega \propto q^{2}$ (Ref.\ \onlinecite{34})
in a free standing film which are characterized by a linear
dependence of phonon state density on $\omega$~: $F(\omega) \propto
\omega$. Since $q_{T} = 2kT/\hbar s_{l,t}$, for Au films 10 nm
thick the above condition is realized at temperatures below 3 K. At
higher temperatures the phonon 
spectrum of free standing film is also quantized \cite{35}. Thus, one 
can expect a 2D behavior of phonons due to the size 
quantization in a free standing film or in a film with weak 
adhesion to the substrate. For the film on substrate only those phonons 
are prone to quantization which have undergone complete 
internal reflection from the metal-substrate boundary, i.e. fly up 
to the boundary at an angle larger than the critical one $\theta_{cr}$
($\sin \theta_{cr} = s_{l,t}/s_{l,t}^{\ast}$, the angle is measured 
from the normal to the boundary and  $\ast$ refers to the substrate). 
For the considered gold-sapphire system $\theta_{cr}^{l} = 17^{\circ}$
and $\theta_{cr}^{t} = 10^{\circ}$. The phonons within the 
critical angle accomplish acoustic metal-substrate coupling. 
\par
We should bear in mind that in the studied real film-substrate 
systems the acoustic coupling is far from being weak. Therefore, 
the results for free films cannot be applied in our case in full 
measure.
\par
The interpretation of the problem proposed in Ref.\ \onlinecite{36}
takes into account the fact that the film is on a substrate. Concurrent 
with the above mentioned waves, we suppose that Love's waves may occur 
in the film-substrate system. The Love's waves are surface waves 
whose displacement vector is parallel to the surface and 
perpendicular to the direction of propagation. Solution of the 
problem for the Love's waves in a film-substrate system results in 
an uncommon dispersion relation. Thus, when the conditions 
$s_{l,t} \ll s_{l,t}^{\ast}$ and $d^{\ast} \ll d$ ($d$ being the 
medium density) are met, the dispersion relation for the Love's waves is 
of the form $\omega \propto q_{T}^{1/2}$ (Ref.\ \onlinecite{37})
similar to that for 2D plasmons. These conditions are well 
fulfilled for the gold-sapphire system. The 1D Love's 
wave which obeyed the above dispersion relation, gives rise to a 
linear frequency spectrum \cite{36}, which is possibly responsible for 
the observed behavior of the temperature dependence of $\tau_{ep}^{-1}$
for the gold films on sapphire substrates.

\subsection{The effect of disordering on $\tau_{so}$ in thin films}

\par
Now we discuss the behavior of $\tau_{so}$ on disordering in the
temperature region where $\tau_{s} \gg \tau_{\varphi 0},
\tau_{so}$. The values of $\tau_{so}$ was found to increase with 
disorder and the increase in $R_{\Box}$ (see the Table). This 
result is quite unexpected because it is commonly assumed that the 
increase in the frequency of elastic scattering processes
$\tau^{-1}$ also involves an increase in the frequency of
spin-orbit processes $\tau_{so}^{-1}$. The ratio of the above
frequencies, $\tau_{so}^{-1}/\tau^{-1}$, characterizes the
probability of spin-orbit process on elastic electron scattering.
An approximate estimation of this quantity can be given by the
following relationship for metals with inversion symmetry of
crystal lattice \cite{38}:

\begin{equation}
\tau_{so}^{-1}/\tau^{-1} \sim (\alpha Z_{i})^{4}, 
\label{nine}
\end{equation}

  if the spin-orbit interaction is of importance in the field of a 
heavy impurity, or 

\begin{equation}
\tau_{so}^{-1}/\tau^{-1} \sim (\alpha Z)^{4} 
\label{ten}
\end{equation}

  if the spin-orbit interaction is essential in the matrix (on
scattering by a light impurity). Here $\alpha = e^{2}/\hbar c
\simeq 1/137$ is the fine structure constant, $Z_{i}$ and $Z$ are
the atomic numbers of the impurity and the matrix (host metal),
respectively.

\par
The dependence of $\tau_{so}^{-1}/\tau^{-1}$ on $Z$ defined by Eq.\
(\ref{ten}) was verified for films of ten different metals in Ref.\
\onlinecite{39}. The verification was made for $\tau_{so}$
determined from the Knight shift, critical magnetic field and
tunnel conductivity of superconductors. The time of electron flight
from surface to surface, $\tau^{sf} = L/v_{F}$, that is the time
between two acts of surface scattering, was used as $\tau$. The
dependence of $\tau^{sf}/\tau_{so}$ on $Z$ was truly close to
$Z^{4}$, and the values of $\epsilon = \tau^{sf}/\tau_{so}$
corresponded to the probability of surface scattering followed by a
spin flip appeared to be rather high ($5\times 10^{-1}$--$5\times
10^{-3}$) for heavy metals (Pb,Sn,Ga,Cu) compared to the expected
values for bulk scattering (approximately by a factor of three). In
Refs.\ \onlinecite{40,41} the values of $\tau_{so}$ for different
degrees of disordering were determined from the effects of weak
localization in Mg, Ag and Al films and the validity of the
relation $\tau_{so}^{-1}/\tau^{-1} = \epsilon = const$ was
confirmed.

\par
One would think that the uncommon behavior of $\tau_{so}$ with
disordering, observed by us in the Au films, could be attributed to
the modification of the nature of main scatterers in the film after
irradiation with Ar ions: the number of scatterers and, hence, the
frequency of elastic scattering processes, $\tau^{-1}$, after the
irradiation increases while the proportion of scatterers with
strong spin-orbit interaction decreases. This could have been due
to the implanted Ar ions if the latter had a weak spin-orbit
interaction compared to the matrix atoms. However, it seems to us
that the high concentration of Ar ions in the film is unlikely
because for their average energy of 2 keV their free path within
the irradiated film is more than the film thickness (a significant
part of ions with high energy pierces the film and penetrates the
SiO sublayer). Moreover, the film heating on irradiation (up to $
\simeq 500$--$600^{\circ}$C) results in that a large part of the
implanted Ar ions escapes it. By the data given in Ref.\
\onlinecite{42}, the maximum Ar concentration in irradiated films
of noble metals is no more than 1\%. For significant influence of
Ar impurities on $\tau_{so}$, it is necessary that the Ar
concentration should be higher at least by the order of magnitude.

\par
Let us assume that the Ar irradiation results in an increase 
in the concentration of radiation defects of the film structure and 
leads to enhancement of diffusion behavior of electron motion in 
the film. If the dominant mechanism of elastic scattering in the 
perfect film is a surface scattering, it turns out that the 
enhancement of disordering causes a bulk scattering to be dominant. 
As in Ref.\ \onlinecite{43}, the frequency $\tau_{so}$ can be 
represented as a sum of contributions from bulk and surface
scattering to spin-orbit relaxation:

\begin{equation}
\frac{1}{\tau_{so}} = \frac{1}{\tau_{so}^{b}} +
\frac{1}{\tau_{so}^{sf}} = 
\frac{\epsilon^{b}}{\tau_{b}} + \frac{\epsilon^{sf}}{\tau^{sf}}.
\label{11}
\end{equation}

It is quite possible that the probability of spin flip under 
electron scattering by surface is much higher than under scattering 
by impurity in the bulk, i.e. $\epsilon^{sf} \gg \epsilon^{b}$.
This suggestion is supported by the experimental results presented
in Ref.\ \onlinecite{43}, where the dependence of $\tau_{so}^{-1}$
on Mg film thickness was studied. The same result was obtained for 
films of different metals in Ref.\ \onlinecite{39} and also for 
Bi films in Ref.\ \onlinecite{44} where the localization correction
to magnetoresistance was studied in parallel magnetic field in
conditions where the quantum size effect manifests itself. 

\par
In view of the above inequality, the processes of spin-orbit
scattering by surface are to be dominant in perfect films where the
second term prevails. With enhancing disorder, the contribution of
these processes decreases because of the increase in the time,
$\tau^{sf}$, of diffusion electron motion from surface to surface.
The contribution of spin-orbit scattering by impurities in the bulk
material having the probability $\epsilon^{b} \ll \epsilon^{sf}$
becomes appreciable. This increases the time $\tau_{so}$. We can
illustrate this by a simple transformation of Eq.\ (\ref{11}). Let
us assume, that $\tau^{b} \approx l/v_{F}$, and the time of
diffusion from one surface to the other is $\tau^{sf} \approx
L^{2}/D$, and $D \approx v_{F}l$.  Then Eq.\ (\ref{11}) becomes 

\begin{equation}
\frac{1}{\tau_{so}} = \frac{\epsilon^{b}v_{F}}{l} + 
\frac{\epsilon^{sf}v_{F}l}{L^{2}}
\label{12}
\end{equation}

or

\begin{equation}
\frac{l}{v_{F}}\frac{1}{\tau_{so}} = 
\epsilon^{b} + \epsilon^{sf}\frac{l^{2}}{L^{2}}
\label{13}
\end{equation}

The experimental points are shown in Fig.~\ref{Fig.6} in the 
$(l/v_{F})/\tau_{so}$ vs. $l^{2}$ coordinates. The straight line in 
Fig.~\ref{Fig.6} obtained by the least-square technique has 
$\epsilon^{b} = 4\times 10^{-4}$  and $\epsilon^{sf} = 2\times 10^{-2}$, 
which supports the assumption that $\epsilon^{sf} \gg \epsilon^{b}$. 
Thus, with the high probability of spin flip during the electron
scattering at the surface, the enhanced disorder in the film
reduces the frequency of collisions with the surface and is
responsible for the decrease in the spin-orbit scattering rate,
observed with decreasing of electron mean free path. 

\section{Conclusions}

\par
The irradiation of Au films with 3.5 keV Ar ions has made it 
possible to obtain samples with the crystal lattice disorder in a 
wide range. The experiments resulted in the previously unknown 
influence of disorder on the processes of phase and spin-orbit 
relaxation of electrons.
\par 
1. It is clearly demonstrated that progressing disorder enhances 
the electron-electron scattering and weakens the the electron-phonon 
scattering (Fig.~\ref{Fig.4}).
\par
2. For a particular degree of disordering, the frequency of 
electron-phonon scattering begins to decrease directly with the 
electron mean free path and the exponent $p$ in temperature 
dependence $\tau_{ep}^{-1} \propto T^{p}$  somewhat increases compared 
to $p = 2$ for weakly disordered films (see the Table). 
The variation in $\tau_{ep}$  on 
disordering may be attributed to the pure-dirty limit transition.
\par
3. Attention is attracted to the fact that the reduced values of 
exponents $p$ in the "pure" limits (cf. the theoretical predictions 
for 3D metals) are related to the surface waves of Love-type,
which present in the film-substrate 
system. These waves obey an unusual dispersion law and have a 
linear frequency spectrum.
\par
4. The abnormal decrease in the rate of spin-orbit scattering with 
enhancing disorder is treated assuming that the probability of spin 
flip is higher under electron scattering by surface than by 
impurities in the bulk.
\acknowledgments              
We are very grateful to V.V.Bobkov and V.I.Glushko for Ar-ion 
irradiation of Au films and electron microscopy study, 
A.I.Kopeliovich, V.A.Shklovsky and E.S.Syrkin for their helpful 
discussion of the results and E.Yu.Kopeichenko for his assistance 
in the processing of the experimental data. Our research work has 
been partially supported by the International Science Foundation 
(grant U2G000).

\begin{figure}
\caption{Quantum correction to Au film conductivity versus 
perpendicular magnetic field for the sample with $R_{\Box}  = 
14.7$~$\Omega$ at different temperatures. From top to bottom $T$ = 
0.56, 4.5, 8, 16, 25 K. The circles are experimental results and 
solid lines are the theoretically calculated quantum corrections 
for corresponding temperatures.
}
\label{Fig.1}
\end{figure}

\begin{figure}
\caption{Temperature dependences of $D\tau_{\varphi}$ for the samples with 
$R_{\Box} \simeq 437$~$\Omega$~ (1) and $R_{\Box} \simeq 85$~$\Omega$~ (2). 
Insert: dependence of $1/\tau_{\varphi}T$ on $T$; - - - - $\tau_{ee}$, 
calculated by Eq.(4.2)}
\label{Fig.2}
\end{figure}

\begin{figure}
\caption{Dependence of electron phase relaxation time upon $R_{\Box}$ at 
various temperatures T, K: $\bigcirc$ -- 4, $\Box$ -- 5, $\triangle$ -- 20, 
$\Diamond$ -- 30. The dashed line is 
only a guide to the eye.}
\label{Fig.3}
\end{figure}

\begin{figure}
\caption{Temperature dependence of Au film resistance. Filled 
circles correspond to experimental data, while open ones represent 
theoretical values found from Eq.(3.2), using the $\tau_{\varphi}$  
values calculated from quantum corrections to conductivity in magnetic 
field.}
\label{Fig.4}
\end{figure}

\begin{figure}
\caption{The electron-phonon relaxation time versus electron 
mean free path for different Au films at $T = 20$~K. The dashed line is 
only a guide to the eye.}
\label{Fig.5}
\end{figure}

\begin{figure}
\caption{Fitting the experimental results by Eq.(4.11).}
\label{Fig.6}
\end{figure}

\newpage

\rightline{\bf Table}
\begin{center}
\begin{tabular}{|p{2.5cm}|p{2.5cm}|p{2.5cm}|p{4cm}|}
\hline 
\multicolumn{1}{|c|}{\bf $R_{\Box} \ (\Omega)$}  
& \multicolumn{1}{|c|}{\bf $T_{tr} \ (K)$}  
& \multicolumn{1}{|c|}{\bf \ $p$ }  
& \multicolumn{1}{|c|}{\bf $\tau_{so}$ (s)}  \\
\hline
10.6  &    3.0 &  $2.0^{+}_{-} 0.1 $ & $3.8 \times 10^{-13}$ \\
14.6  &    4.3 &  $2.0^{+}_{-} 0.1 $ & $5.3 \times 10^{-13}$ \\
14.7  &    4.3 &  $2.0^{+}_{-} 0.1 $ & $6.2 \times 10^{-13}$ \\
20.4  &    6.0 &  $2.0^{+}_{-} 0.1 $ & $5   \times 10^{-13}$ \\
44.4  &    11  &  $1.9^{+}_{-} 0.1 $ & $8.5 \times 10^{-13}$ \\
61.4  &    18  &  $2.0^{+}_{-} 0.1 $ & $1   \times 10^{-12}$ \\
85    &    25  &  $2.2^{+}_{-} 0.1 $ & $3   \times 10^{-12}$ \\
143   &    37  &  $2.3^{+}_{-} 0.1 $ & $4.8 \times 10^{-12}$ \\
161   &    43  &  $2.55^{+}_{-} 0.1$ & $6.1 \times 10^{-12}$ \\
437   &   103  &  $2.8^{+}_{-} 0.15$ & $1.2 \times 10^{-11}$ \\
\hline
\end{tabular}
\end{center}
\end{document}